\documentclass[apj, 11pt]{emulateapj}
\usepackage{epsfig}
\usepackage{color}
\shorttitle{Shock Turbulence Interaction}

\begin{document}

\title{On the Amplification of Magnetic Field by a Supernova Blast Shock Wave in a Turbulent Medium}

\author{Fan Guo \altaffilmark{1,2}, Shengtai Li \altaffilmark{1}, Hui Li \altaffilmark{1},
Joe Giacalone \altaffilmark{2}, J. R. Jokipii \altaffilmark{2} and David Li \altaffilmark{1}}

\altaffiltext{1}{Theoretical Division, Los Alamos National Laboratory, Los
Alamos, NM 87545}

\altaffiltext{2}{Department of Planetary Sciences and Lunar and Planetary
Laboratory, University of Arizona, 1629 E. University Blvd., Tucson, AZ 85721}

\begin{abstract}
We have performed extensive two-dimensional magnetohydrodynamic simulations to study the amplification of magnetic fields when a supernova blast wave propagates into a turbulent interstellar plasma. The blast wave is driven by injecting high pressure in the simulation domain. The interstellar magnetic field can be amplified by two different processes, occurring in different regions. One is facilitated by the fluid vorticity generated by the ``rippled" shock front interacting with the background turbulence. The resulting turbulent flow keeps amplifying the magnetic field, consistent with earlier work \citep{Giacalone2007}. The other process is facilitated by the growth of the Rayleigh-Taylor instability at the contact discontinuity between the ejecta and the shocked medium. This can efficiently amplify the magnetic field and tends to produce the highest magnetic field. We investigate the dependence of the amplification on numerical parameters such as grid-cell size and on various physical parameters. We show the magnetic field has a characteristic radial profile that the downstream magnetic field gets progressively stronger away from the shock. This is because the downstream magnetic field needs a finite time to reach the efficient amplification, and will get further amplified in the Rayleigh-Taylor region. In our simulation we do not observe a systematic strong magnetic field within a small distance to the shock. This indicates that if the magnetic-field amplification in supernova remnants indeed occurs near the shock front, other processes such as three-dimensional instabilities, plasma kinetics and/or cosmic ray effect may need to be considered to explain the strong magnetic field in supernova remnants.
\end{abstract}


\section{Introduction}
Powerful shocks associated with supernova remnants (hereafter SNRs) sweeping
through the interstellar medium (ISM) are remarkable high-energy phenomena in
astrophysics. It is widely believed that these high-Mach number shocks are
the sources of galactic cosmic rays with energies up to at least $10^{15}$ eV.
SNRs are also sources of strong radio and/or X-ray emissions. In these high-energy
processes, the magnetic field is of great importance. Moreover, it provides information
on energetic charged particles, which are presumably accelerated by the supernova shocks.

The ISM is known to be turbulent. Measurements of the ISM radio-wave scintillation
have established the existence of large-scale density turbulence which has a
Kolmogorov-like power spectrum spanning more than ten decades of spatial scale
with an outer scale of several parsecs
\citep[e.g.,][]{Lee1976ApJ,Armstrong1981,Rickett1990,Armstrong1995,Minter1996}.
This has been called ``the big power law in the sky." \citep[see,][]{Spangler2007}
The galactic magnetic field is observed to be a few micro-Gauss and has
uniform and fluctuating components that are roughly in equipartition
\citep[e.g.,][]{Beck1996,Minter1996,Han2004}. The turbulent magnetic field can
interact with the shock waves, distorting their surfaces leading to
shock ripples \citep{Neugebauer2005} and the enhanced downstream magnetic
fluctuations \citep{Lu2009}. It is also important
for efficient particle acceleration \citep{Giacalone2005,Jokipii2007,Guo2010a}.
Turbulence in the upstream medium has also been considered \citep[][]{Balsara2001}
to explain the irregular and patchy emission morphology observed in SNRs
\citep[e.g.,][]{Anderson1996}.

Recently, it has been inferred from observations that the magnetic field
in young SNRs is strongly enhanced to a magnitude much greater than the
compression given by the shock jump condition. For example, by assuming that the so-called X-ray
``thin rims" seen in several young SNRs \citep[e.g.,][]{Bamba2005} are caused by shock accelerated
electrons rapidly losing energy in strong magnetic field through synchrotron radiation, the
associated magnetic field may be more than $50-100\mu G$ in order to explain the thickness
of the ``thin rims" \citep{Berezhko2003,Volk2005,Ballet2006,Parizot2006}.
For SNR shocks with higher shock speeds propagating in more
inhomogenous media such as Cas A and Tycho, the downstream magnetic
fields  are inferred to be more enhanced. ``Thin rims" are also seen
in radio emissions \citep{Reynoso1997}, which cannot be explained by the electrons losing energy in
strong magnetic field. This indicates some other mechanism, e.g.,
decay of magnetic fluctuations may need to be considered \citep{Pohl2005}. Further downstream of
the shock, the magnetic field may possibly be even higher than the region right
behind the shock \citep{Vink2003}.
The morphology of X-ray emission in SNRs shows filamentary structure and
rapid time variation, which indicates that the magnetic field could be as high
as $1 mG$ \citep{Uchiyama2007}, over two orders of magnitude higher than the background fluid.
It should be pointed out this rapid variation of synchrotron emission could be well reproduced in the case
of strong magnetic fluctuations \citep{Bykov2008}. Since the effect of turbulence is not fully
understood at this point, the actually magnetic-field amplification factors in young SNRs
remain uncertain.

\citet{Bell2001} and \citet{Bell2004} proposed that cosmic rays accelerated by
the SNR forward shock waves provide a current that leads to an
instability that can amplify the magnetic field close to the shock front.
Numerical simulations show the evidence of this instability, however, it is found
to easily saturate
and the amplification factor may be limited \citep[e.g.,][]{Riquelme2009}.
Recently, \citet{Giacalone2007} proposed an alternative mechanism, in
which the interaction between the warped shock front and large-scale
density fluctuations produces fluid vorticity downstream of strong
shocks. That fluid vorticity can stretch, distort and amplify the
magnetic field. The magnetic-field amplification in this mechanism
relies on the dynamics of a magnetized fluid rather than the cosmic-ray
kinetic physics. It is interesting to note, however,
\citet{Balsara2001} performed three-dimensional MHD blast
wave simulation with moderate resolution which does not show magnetic-field
enhancement larger than $50 \mu G$, whereas two-dimensional
Cartesian geometry simulations with high numerical resolution give strong
amplification \citep{Giacalone2007,Inoue2009} with maximum values
larger than $100\mu G$. This discrepancy warrants further investigation.

In addition, a critical constraint from several observed ``thin rims" is that
the required magnetic-field amplification should occur within
a narrow distance $\lesssim 0.01$ pc of the supernova shocks \citep[for a review, see][]{Reynolds2011}.
This is supported by some coincidences of X-ray ``thin rims" and shock locations inferred from H$\alpha$ observations \citep{Winkler2003} and radio polarimetry \citep{Gotthelf2001}. This places an
important constraint on various field amplification models, although it should be noted that
there is no one-to-one correspondence between X-ray ``thin rims" and inferred shock locations. To our knowledge,
this constraint has not been taken into account previously in comparing models
and simulations with observations.

In this study, we perform a series of two-dimensional ideal MHD
simulations with high spatial resolution
to study strong supernova blast
shock waves propagating into the ISM containing pre-specified
large-scale density and magnetic fluctuations. We investigate how the
amplification depends on a variety of parameters, including the explosion energy,
the level of background turbulence, and the numerical resolution, etc.
The paper is organized as follows. In Section 2, we describe our numerical
model and simulation setup. We present the simulation results in Section 3, and
in Section 4 we summarize and discuss our results.

\section{Basic Consideration and Numerical Model}

We have performed a series of two-dimensional  ideal MHD
simulations with high-order numerical schemes.
The ideal MHD equation can be expressed as:
\begin{eqnarray}
\partial_t \rho + \nabla \cdot (\rho \textbf{u}) &=& 0 \\
\partial_t ( \rho \textbf{u}) + \nabla \cdot (\rho \textbf{u} \textbf{u}
+ (p+ \frac{B^2}{8\pi})\vec{\vec{\textbf{I}}}  - \frac{\textbf{BB}}{4\pi}   )  &=& 0 \\
\partial_t (E) + \nabla \cdot ((E+p+\frac{B^2}{8\pi})\textbf{u} -
\frac{\textbf{B}(\textbf{u}\cdot \textbf{B})}{4\pi})    &=& 0 \\
\partial_t \textbf{B} - \nabla \cdot (\textbf{uB} - \textbf{Bu}) &=& 0
\end{eqnarray}
where $\rho$ is the plasma density, $\textbf{u}$ is the fluid velocity,
$\textbf{B}$ is the magnetic field, $p$ is the gas pressure, and $E$ is the
total energy density, which is defined as:
\begin{equation}
E = \frac{p}{\gamma - 1} + \frac{1}{2}\rho u^2 + \frac{1}{8\pi}B^2~~.
\end{equation}

We use a central finite-volume scheme on overlapping grid cells \citep{Liu2007}
to solve the ideal MHD equations. In particular, a high-order
divergence-free reconstruction for the magnetic field that uses the
face-centered values has been employed \citep{Li2008,Li2010}. The magnetic field
is advanced with a high-order
constrained transport (CT) scheme to preserve the divergence-free condition to
machine round-off error. The overlapping cells are natural to be used to
calculate the electric-field flux without a spatial-averaging procedure, and
hence we can achieve the higher-order accuracy (3rd-order or 4th-order).
The central schemes do not need time-consuming characteristic
decompositions, and are easy to code and be combined with un-split discretization
of the source and parabolic terms. The overlapping cell representation of the
solutions is also used to develop more compact reconstruction and less
dissipative schemes. For solutions that contain discontinuities, e.g., shocks and/or
contact discontinuities, we apply a non-oscillatory hierarchical reconstruction (HR)
to remove the spurious oscillations and achieve high resolution near the discontinuities.
The HR limiting we have used \citep{Li2008,Li2010} requires information only from the
nearest neighbor cells and it does not require characteristic decomposition. The
numerical dissipation introduced by the HR limiting is enough to damp out all the
artificial oscillations near the shocks and no extra artificial viscosity is needed.
To further improve the computational efficiency and reduce the numerical dissipation
for the smooth flow, we develop a shock detector to flag the cells near the shock
and perform HR only to those cells. The details of the whole algorithm have been documented
in \citep{Li2008,Li2010}. Our
method has been verified to achieve the expected order of accuracy and have
very low numerical dissipation. The high-order, low-dissipation, and
divergence-free properties of this method make it an ideal tool for MHD turbulence simulations.

We model the simulation in a two-dimensional Cartesian coordinate ($x, y$)
with uniform grids. The size of the simulation domain corresponds to
$L_x \times L_y = 30 pc \times 30 pc$ in all the simulations. A
supernova blast wave is driven by the initial injection of thermal
pressure and mass in a small circular region at the center of the simulation
box. The physical parameters used in the simulations are listed in Table
$1$. To calculate the injection energy, we must set a volume for the injection region.
We assume a cylinder with radius of $r=0.4 pc$ and length in $z$
direction is $L_z = 0.8 pc$, which gives a volume of $0.402 pc^3$.
Initially, the cylinder region has a high pressure $p_{in}$, and a total mass about $3$ solar masses.
For most cases considered here, we assume that the injected internal energy
in the central region is $1.5 \times 10^{51}$ erg. A grid number of
$n_x \times n_y =4000\times4000$ is typically used for the simulation domain. This initial setup for blast
waves is similar to the simulation made by \citet{Balsara2001}, except that we use
two-dimensional simulations with higher resolutions to study the magnetic-field
evolution in young SNRs.
The density and magnetic field in the background ISM consist of an
average component and a turbulent component. We take the interstellar background
plasma mean number density to be $n_0 = 1$ cm$^{-3}$. The average magnetic field
$B_0$ is along the $y$ direction, whose magnitude is given in Table 1
for different cases. In our model, we assume a constant initial background ISM
pressure $p_0$. Its corresponding temperature $T_0$ for each case is listed
in Table 1. For the fluctuating components, we assume that both
magnetic field and density fluctuations have a two-dimensional
Kolmogorov-like power spectrum:
\begin{equation}
P \propto \frac{1}{1+(kL_c)^{8/3}}~~.
\end{equation}

This is consistent with observations of interstellar turbulence \citep{Armstrong1995,Chepurnov2010}.
The turbulence is generated by summing a large number of discrete wave modes
with random phases \citep{Giacalone1999}. In all the simulation cases the coherence
length of the background turbulence is chosen as $L_c = 3 $ pc. The random component of magnetic
field is given by

\begin{eqnarray}
\delta \textbf{B} (x, y) = \sum^{N_m}_{n=1} &\sqrt{C_B2\pi k_n \Delta k_n
P_B(k_n)} (\sin \theta_n \hat{x} - \cos \theta_n \hat{y})  \nonumber \\
& \times\exp (i \cos \theta_n k_n x + i \sin \theta_n k_n y + i \phi_n)
\end{eqnarray}

where $P_B(k_n)$ is the power for wave mode with wavenumber $k_n$, random
propagation angle $-1<\cos\theta_n<1$ and random phase $0<\phi_n<2\pi$. $C_B$ is
a constant used to normalize the wave amplitude. In this study, the total fluctuating
magnetic field is taken to be $\delta B^2 = B_0^2$ in all cases.

The density fluctuations
satisfy a log-normal probability distribution \citep{Burlaga2000,Giacalone2007}:
\begin{equation}
n(x,y) = n_0\exp(f_0+\delta f)
\end{equation}
in which $f_0$ is a constant used to determine the mean density and $\delta f$
is given by a similar expression as the turbulent part of
the magnetic field

\begin{eqnarray}
\delta f(x, y) =&\sum^{N_m}_{n=1} \sqrt{C_f2\pi k_n \Delta k_n P_f(k_n)}  \nonumber \\
&\times\exp(i \cos \theta_n k_n x + i \sin \theta_n k_n y + i \phi_n)
\end{eqnarray}

In Runs 1 - 4 we consider the effect of the amplitude of upstream density fluctuations.
In Runs 5 and 6 we examine the effect of different value of $B_0$.
In Runs 7 and 8 we examine the effect of different temperatures of the
ISM. We have also simulated two different explosion energies in Run 9
 and Run 10.
In Runs 11 and 12, we examine the effect of the simulation grid-cell size.
Runs 1 and 2 are the representative runs with and without pre-specified background turbulence.

\section{Simulation Results}

We now describe the simulation results on the interaction between the supernova blast wave
and the turbulent ISM, along with the magnetic-field evolution downstream of the supernova
shock. We will first present the results from our representative Run 1. Then we will discuss the
spatial distribution of magnetic-field amplification, followed by the
effects of numerical resolution and
other effects such as the background turbulence and explosion energy.

\subsection{Two Regions of Magnetic-field Amplification}

\begin{figure}
\begin{center}
\includegraphics[width=80mm]{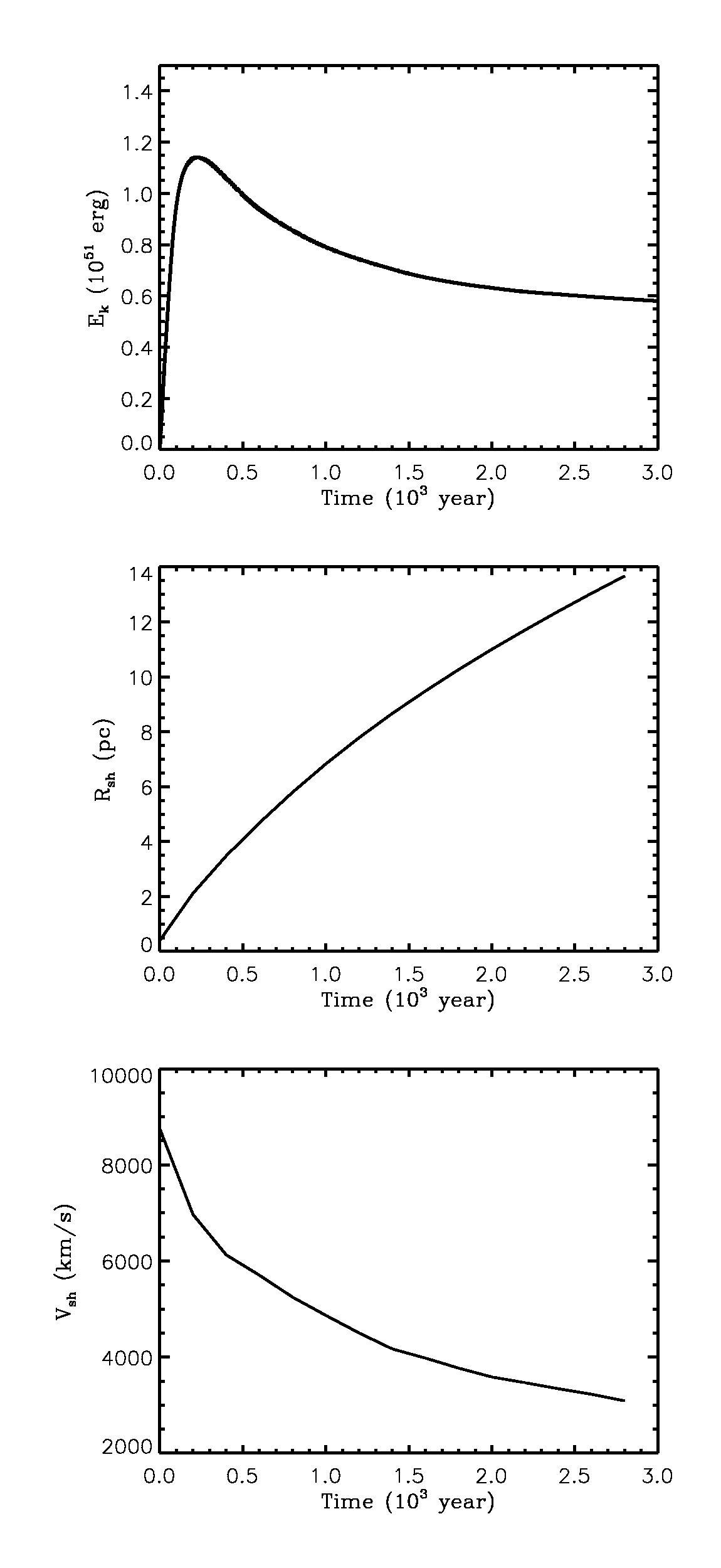}
\caption{The time evolution of (a) the total kinetic energy summed over the simulation domain, (b) the average radius, and (c) the average speed of the shock for Run 1. The initial injected thermal energy is $1.5\times 10^{51}$ ergs.}
\label{fig1}
\end{center}
\end{figure}

Figure \ref{fig1} shows the time evolution of (a) the total kinetic energy summed over the whole simulation domain,
(b) the average radius of the blast wave, and (c) the average speed of
the shock for Run $1$. The average radius is estimated using the area of the high pressure region for
each snapshot of pressure in the simulation, and the average speed of the shock is calculated by taking running differences between
the averaged radii.
In the beginning of the simulation, the region with high density and high pressure
expands and drives a shock propagating into the turbulent medium.
The kinetic energy increases sharply during the expansion and reaches about
$1.15 \times 10^{51}$ erg, i.e., $\sim 77\%$ of the injected explosion energy is converted to
kinetic energy at its peak. After that the swept-up ISM slows
down the ejecta so the kinetic energy decreases slowly. In all the cases we
have simulated, the total energy is conserved within
a degree of $10^{-6}$ during the simulation time. The shock speed
decreases from about 8700 km/s to about 3000 km/s at the end
of the simulation. The corresponding Alfv\'en Mach number changes from $\sim 850$ to $\sim 290$.
The radius of the remnant roughly follows $r \propto t^{0.8}$,
which is consistent with the self-similar solution \citep{Chevalier1982}.
 The size, age and shock speed are
roughly consistent with the observations of young SNRs, though there is a wide spread
in these quantities observationally.
After a few times of the ejected mass being swept by the supernova blast wave, the shock speed is expected to
slow down and settle into an ``Sedov" phase in which $r \propto t^{0.4}$ typically after a few thousand years.
Since in three-dimension the interstellar mass swept by supernova shock is proportional to $r^3$ rather than $r^2$
in our two-dimensional simulation, the shock speed in later times (after several thousand years) is likely overestimated.
In this work we mainly focus on the evolution of young SNRs.

\begin{figure}
\begin{center}
\includegraphics[width=80mm]{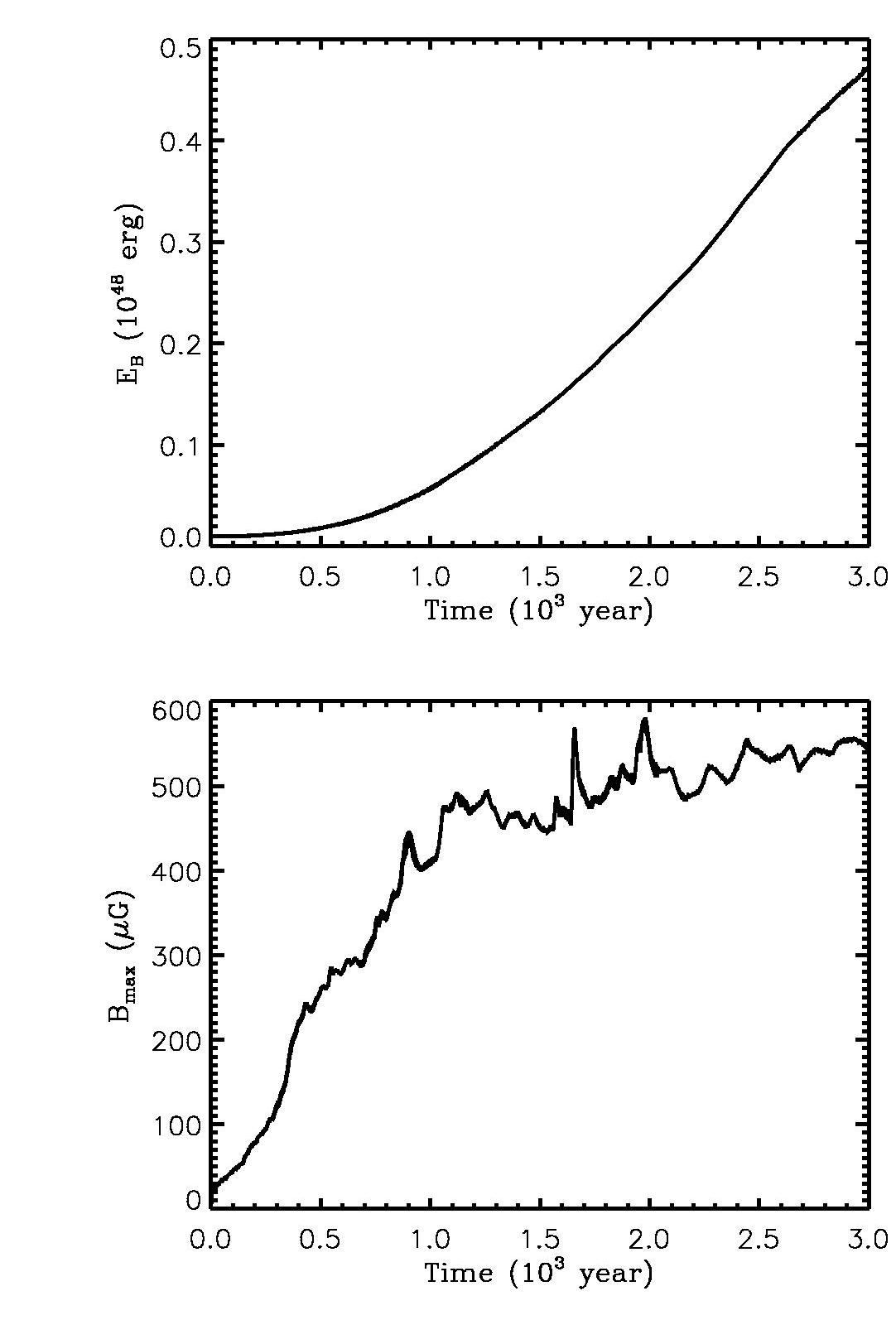}
\caption{The time evolution of (a) the total magnetic-field energy $E_{\rm B}$ and (b) the maximum magnetic-field strength
in the simulation domain $B_{\rm max}$, for Run 1.}
\label{fig2}
\end{center}
\end{figure}

Figure \ref{fig2} $(a)$ and $(b)$ display the total magnetic-field energy and the maximum
magnetic-field strength in the simulation box for Run 1.
It is shown that the magnetic-field energy increases to a level much
larger than the initial magnetic-field energy.
At the end of the simulation, the magnetic-field energy reaches to
$\sim 0.47 \times 10^{48}$ erg. These results are qualitatively consistent
with previous studies using the planar shock waves \citep{Giacalone2007,Inoue2009}.
The maximum magnetic field can rapidly increase to $250\mu G$
within 500 years and finally goes up to about $550 \mu G$ in 3000 years, much larger
than what is expected from the jump condition. But the detailed analysis shows that
the spatial location of the maximum magnetic field is typically \emph{not} near the
immediate downstream region of the shock. We will discuss this result in more detail later.

\begin{figure}
\centering
\begin{tabular}{cc}
\epsfig{file=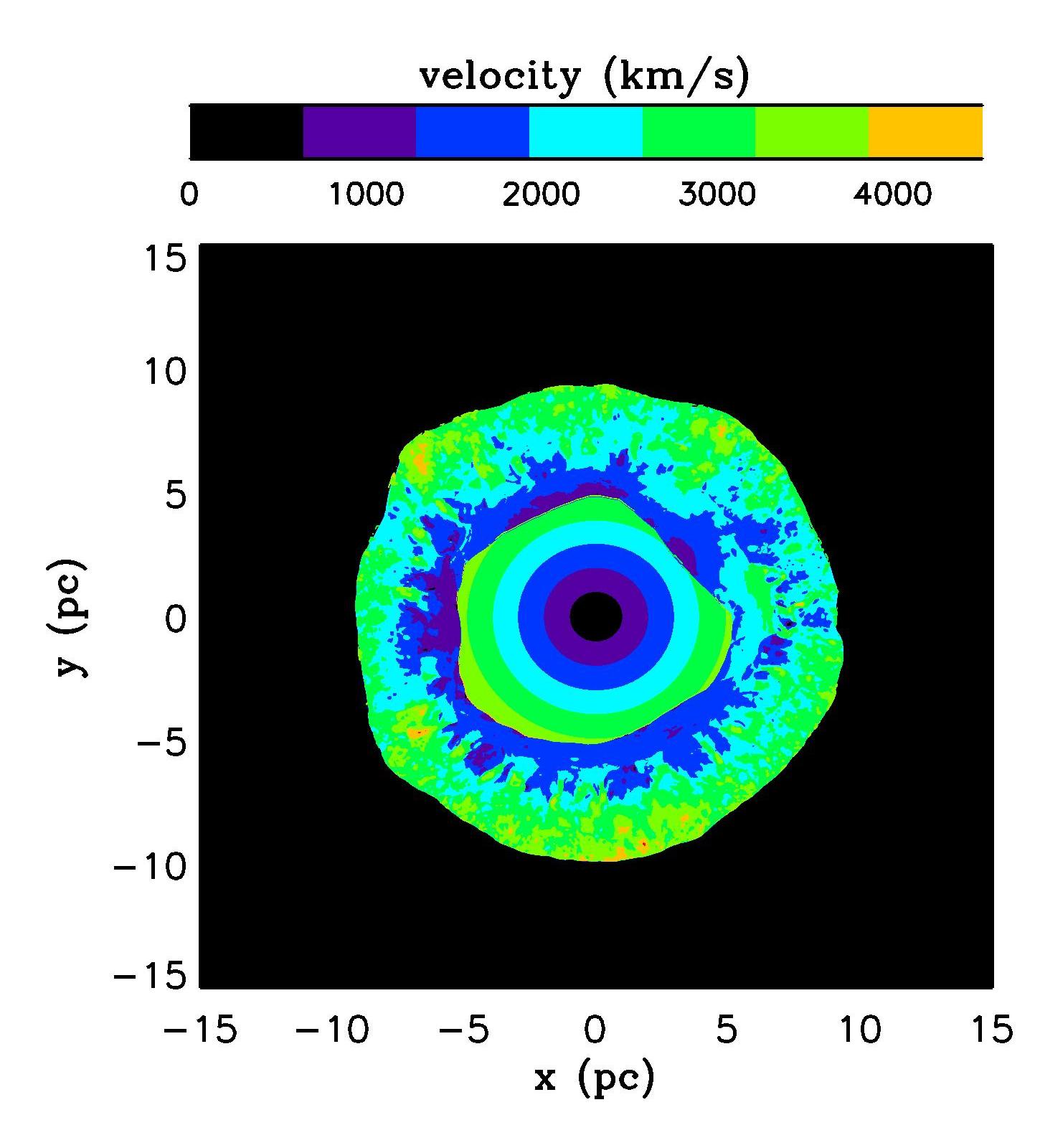,width=0.4\textwidth,clip=} &
\epsfig{file=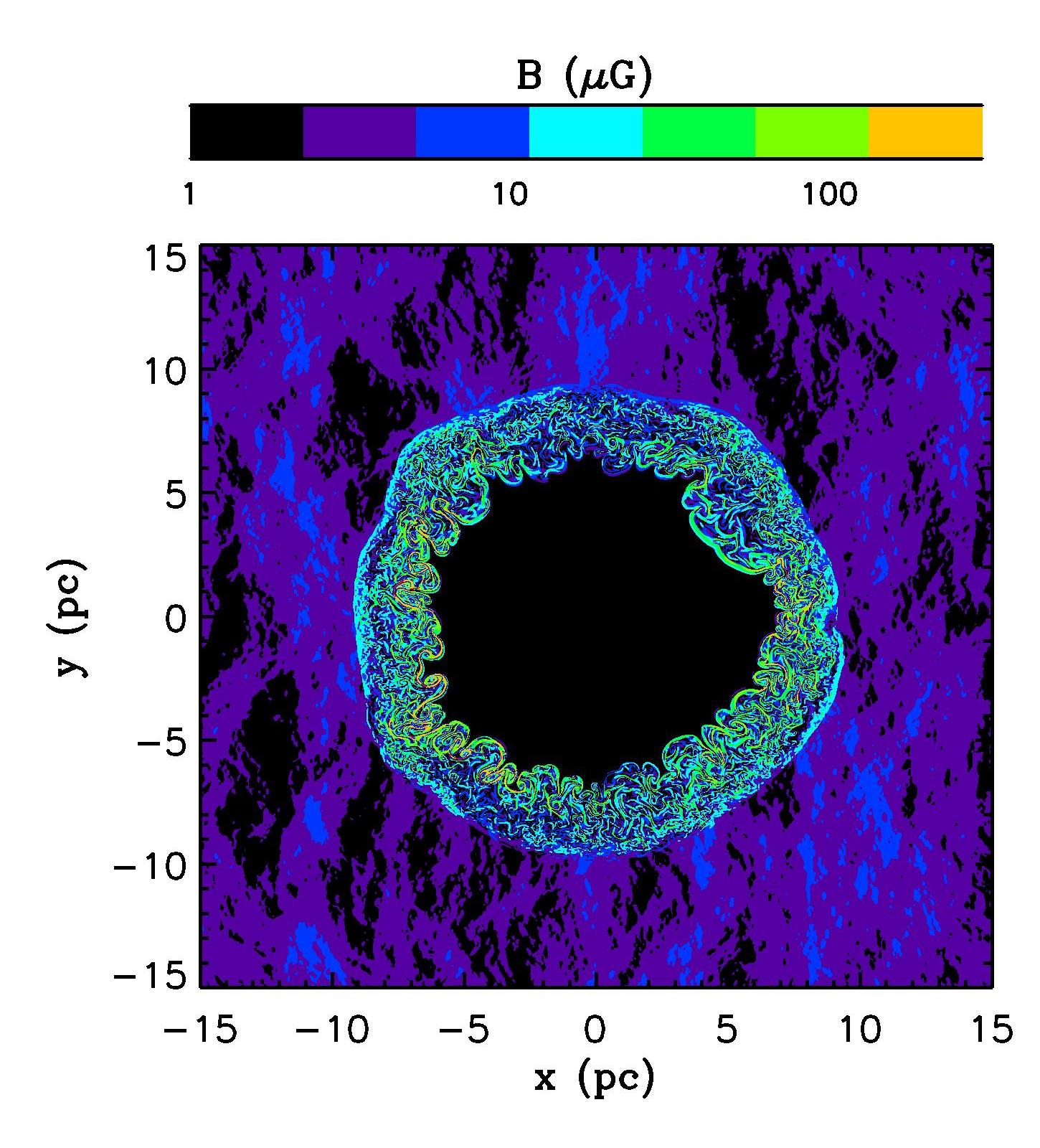,width=0.4\textwidth,clip=} \\
\epsfig{file=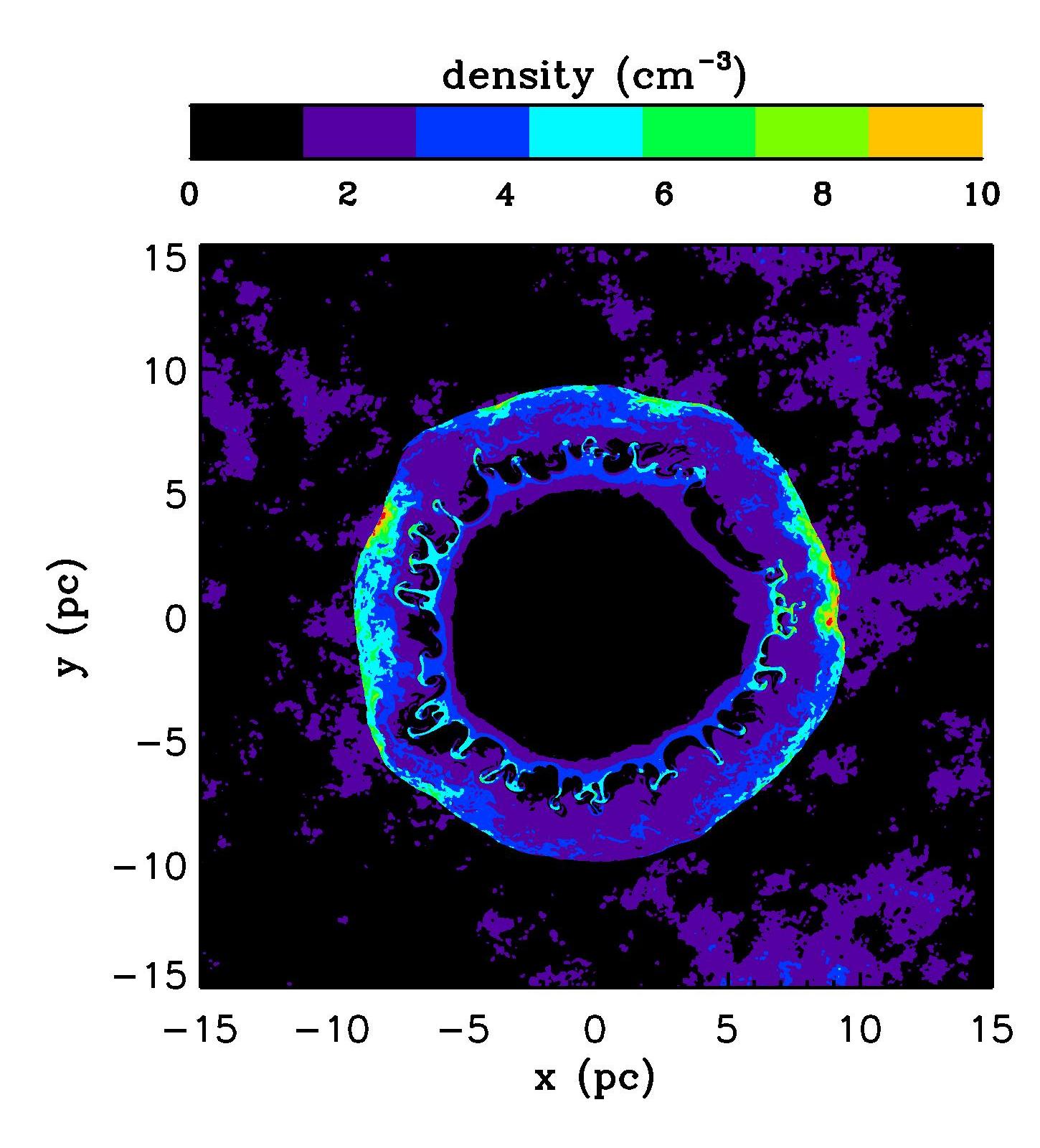,width=0.4\textwidth,clip=} &
\epsfig{file=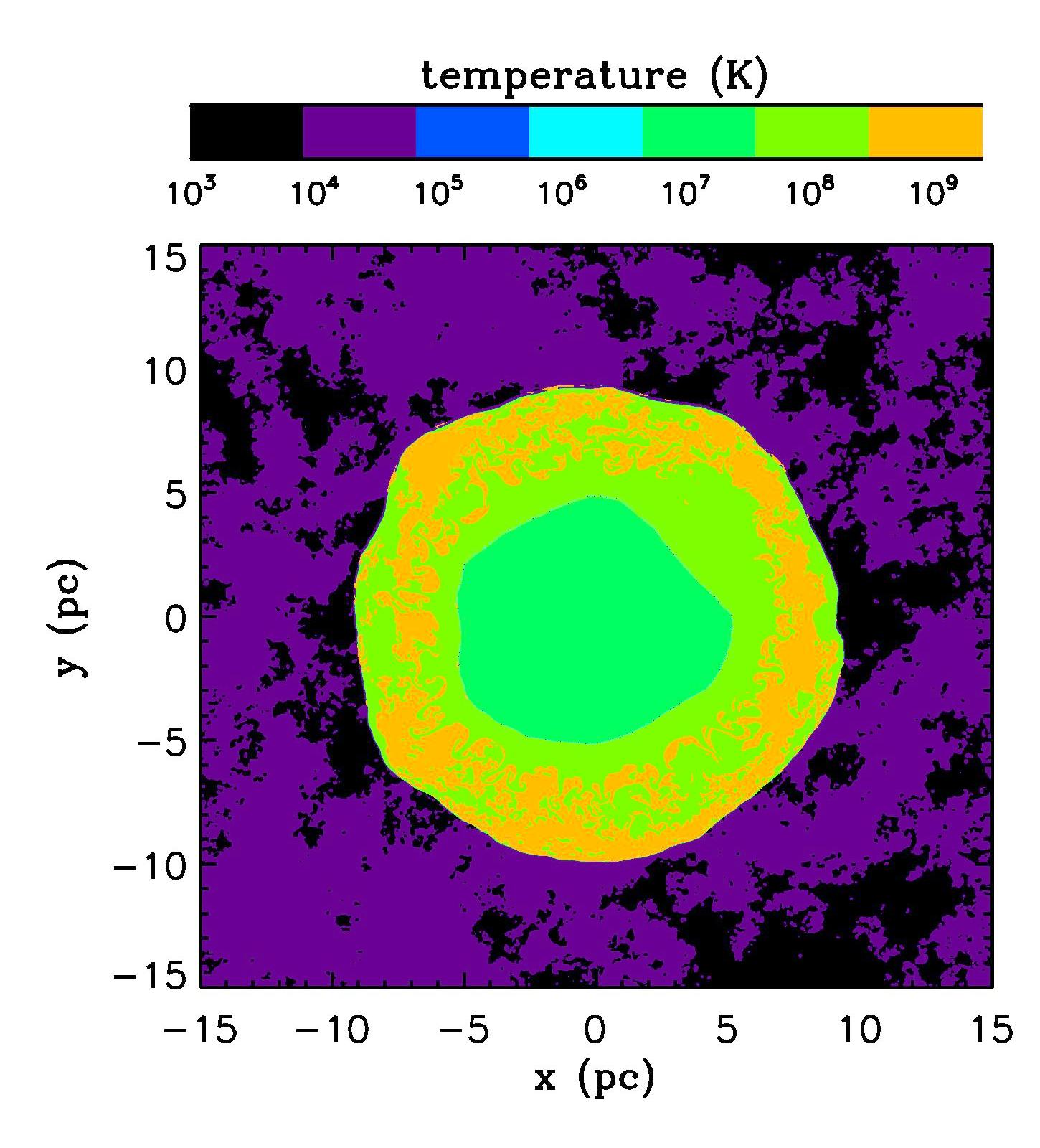,width=0.4\textwidth,clip=}
\end{tabular}
\caption{Spatial distribution of (a) the velocity magnitude, (b) the magnetic-field strength, (c) density, and (d) temperature
for Run 1 at $t =1600$ years. The background ISM turbulence causes the shock front to become rippled.
In addition, the interface between the ejecta and the shocked ISM experiences the Rayleigh-Taylor-like instabilities.}
\label{fig3}
\end{figure}

Figure \ref{fig3} shows the snapshots of (a) the velocity magnitude,
(b) the magnitude of magnetic field, (c) density and (d) temperature
at $t = 1600$ years for Run 1. It is shown that the velocity field of the
blast wave is highly irregular.  The shock surface is rippled
as regions with different densities pass through the shock front \citep{Giacalone2007}.
The flow at the rippled shock transition produces strong transverse and rotational flow downstream of
the shock wave. It can be seen from Figure \ref{fig3} (b) that the magnetic field downstream of the blast wave
is strongly amplified. We find the amplification is closely related to the downstream vorticity
production \citep{Giacalone2007}. This flow patten stretches and distorts
the field lines of force, which leads to a small-scale dynamo process.

In addition, different from the shock amplification, we find that
the magnetic field in the interface region between the ejecta and the shocked medium
is also strongly enhanced by the Rayleigh-Taylor instability (RTI) at the contact
discontinuity \citep[e.g.,][]{Jun1996a,Jun1996b}. It appears that the magnetic
field can be enhanced to $\sim 300 \mu G$ in the region where RTI
is important.  Figure \ref{fig3} (c) shows ``fingers" of enhanced density resulting from the RTI
in the interface region, which is quite different from the mechanism discussed above, which
occurs just behind the shock.
In fact, the magnetic-field amplification by the RTI process contributes almost equally
to the total enhanced magnetic energy and the maximum field strength as shown in
Figure \ref{fig2}.  The nonlinear development of RTI stretches the magnetic
field and causes strong amplification. This process has been studied extensively
\citep[e.g., see the earlier numerical studies by][]{Jun1995,Jun1996a,Jun1996b}.
The RTI amplification process can be studied in more detail when a more
realistic initial ejecta profile are taken into account.
Though beyond the scope of this paper, it will be worthwhile to see if there are possible
observational signatures in the RTI-amplification region.

\subsection{Spatial Dependence of the Magnetic-field Amplification}

As discussed in the Introduction, some observed X-ray ``thin rims" and their coincidence with
 the inferred shock locations have suggested that magnetic field is amplified at or within a short distance to the supernova remnant shock front. It is thus imperative to investigate whether the MHD simulations can reproduce this feature, although to directly compare to X-ray observation would also require the simultaneous computation
of cosmic-ray variation, which is not included in this study.

\begin{figure}
\centering
\begin{tabular}{c}
\hfill
\epsfig{file=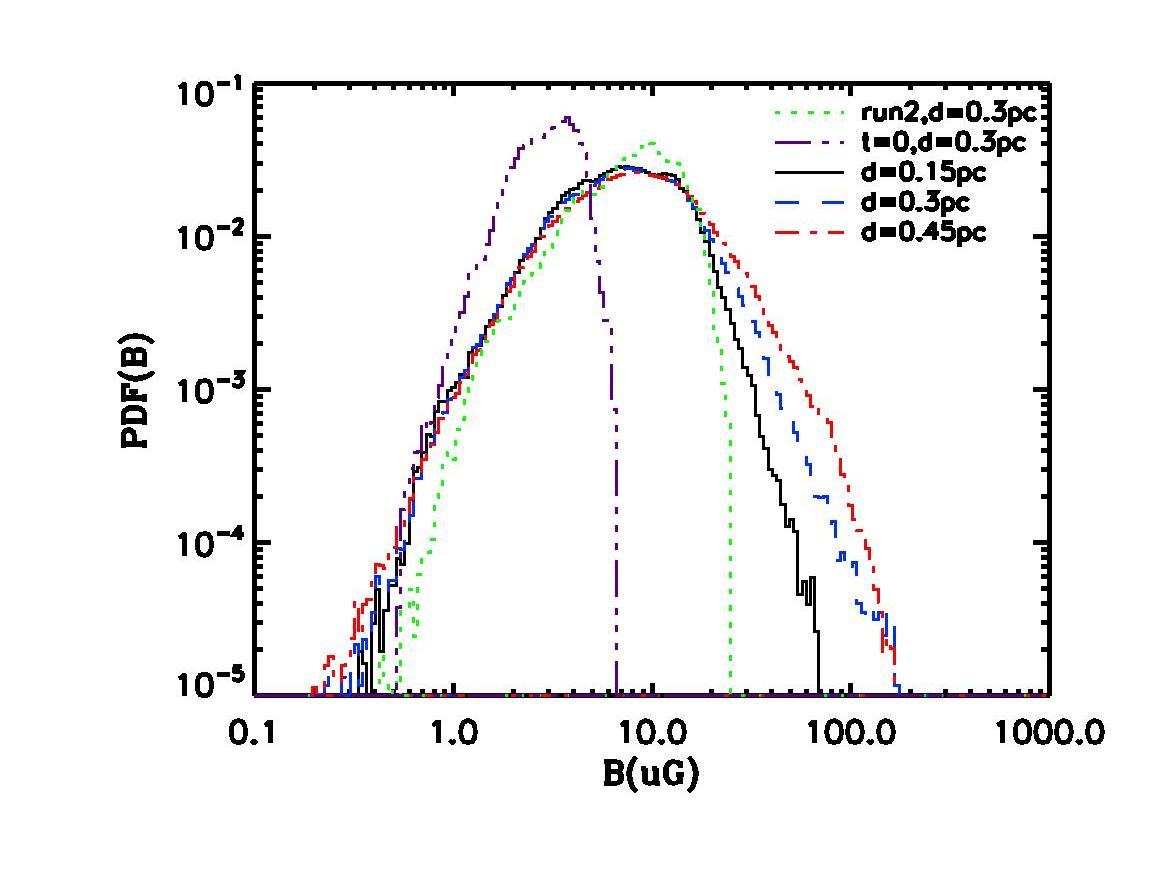,width=60mm,clip=}
\end{tabular}
\caption{The probability distribution function of magnitude
of magnetic field downstream with a distance within $0.15$ pc (black solid line),
$0.3$ pc (blue dashed line) and $0.45$ pc (red dot dashed line) behind the shock front for
Run $12$ at $t = 600$ years. The probability distribution functions for magnetic field
with a distance within $0.3$ pc behind the shock front in
Run $2$ at $t = 600$ years and the initial background magnetic field are also plotted for comparison.}
\label{fig4}
\end{figure}

In Figure \ref{fig4} we present the results at
$t = 600$ years for Run 12,  which has the highest available resolution.
It shows the probability distribution functions (PDFs) of magnetic field, taken within
a thin region of $0.15$ pc, $0.3$ pc
and $0.45$ pc behind the shock front, respectively. In addition, the PDF of the
downstream magnetic field within a distance of $0.3$ pc of the shock front
for Run 2 (without background turbulence) is also plotted for comparison.
The PDFs presented here and in other figures of this paper are normalized by the respective
volume from which the distribution is taken.
It appears that the magnitude of magnetic
field increases with the distance from the shock front.
Specifically, for the rim within $0.15$ pc downstream,
the maximum magnetic field reaches only $70 \mu G$. Also the region with magnetic field
higher than $30 \mu G$ only occupies about $0.8\%$ of the rim.

\begin{figure}
\begin{center}
\hfill
\includegraphics[width=80mm]{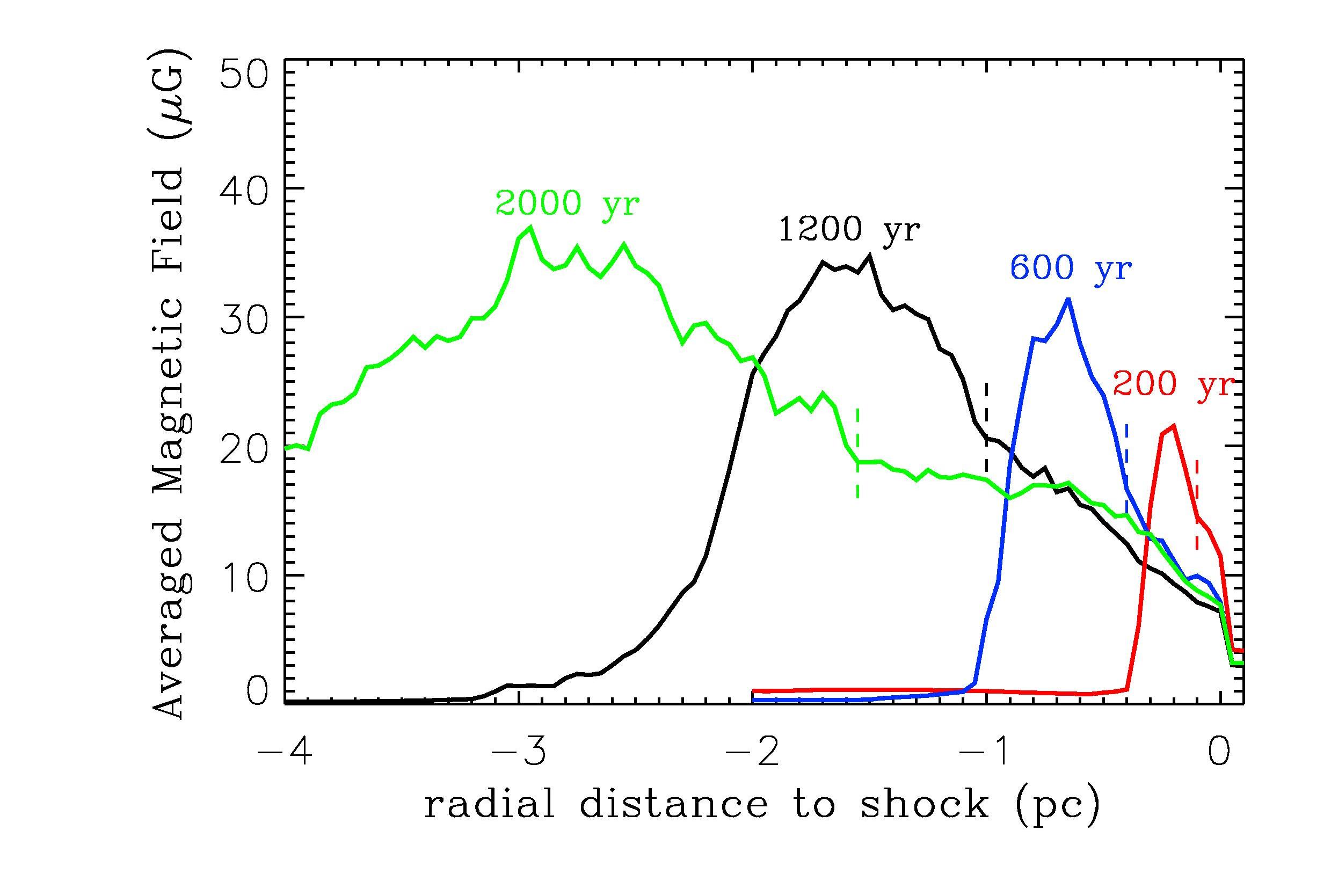}
\caption{The average magnetic-field magnitude as a function of radial distance to the shock front from Run 12 at
different time from 200 years to 2000 years. The dashed lines roughly separate magnetic-field regions dominated by the shock amplification and by Rayleigh-Taylor convective flow.}
\label{fig5}
\end{center}
\end{figure}

To explore further the amplification of the magnetic field in both the shock downstream region
and the Rayleigh-Taylor region, we plot
the average magnitude of magnetic field as a function of radial distance from the shock front
at different times from 200 years to 2000 years for Run 12,
which is shown in Figure \ref{fig5}. Together with the spatial distribution of magnetic field at these times
(not shown here) that is similar to Figure \ref{fig3},
we can separate approximately the two different magnetic-field amplification regions. In Figure \ref{fig5},
the right side of the dashed lines is typically dominated by the shock amplification and the
left is by the Rayleigh-Taylor instability. It is
shown that Rayleigh-Taylor amplification produces much larger magnetic field than those from the
shock amplification. Detailed analysis shows that, locally, the amplified magnetic field in the Rayleigh-Taylor
region can easily reach several hundred micro-gauss.

It can be seen from the downstream magnetic-field evolution that the amplification of magnetic field has
not reached a saturation. This is probably not surprising, given the fast transit time of the SNR shock
and the relatively young age of the SNRs. The gradual increase of the magnetic-field amplitude
further downstream from the shock is consistent with the picture that turbulence has longer time
to amplify the fields. Before the magnetic field reaches Rayleigh-Taylor region, its maximum
value is only about $200-300$ micro-Gauss, much less than the amplified magnetic field in
Rayleigh-Taylor regions.

\subsection{Effects of Numerical Resolution}

\begin{figure}
\begin{center}
\hfill
\includegraphics[width=80mm]{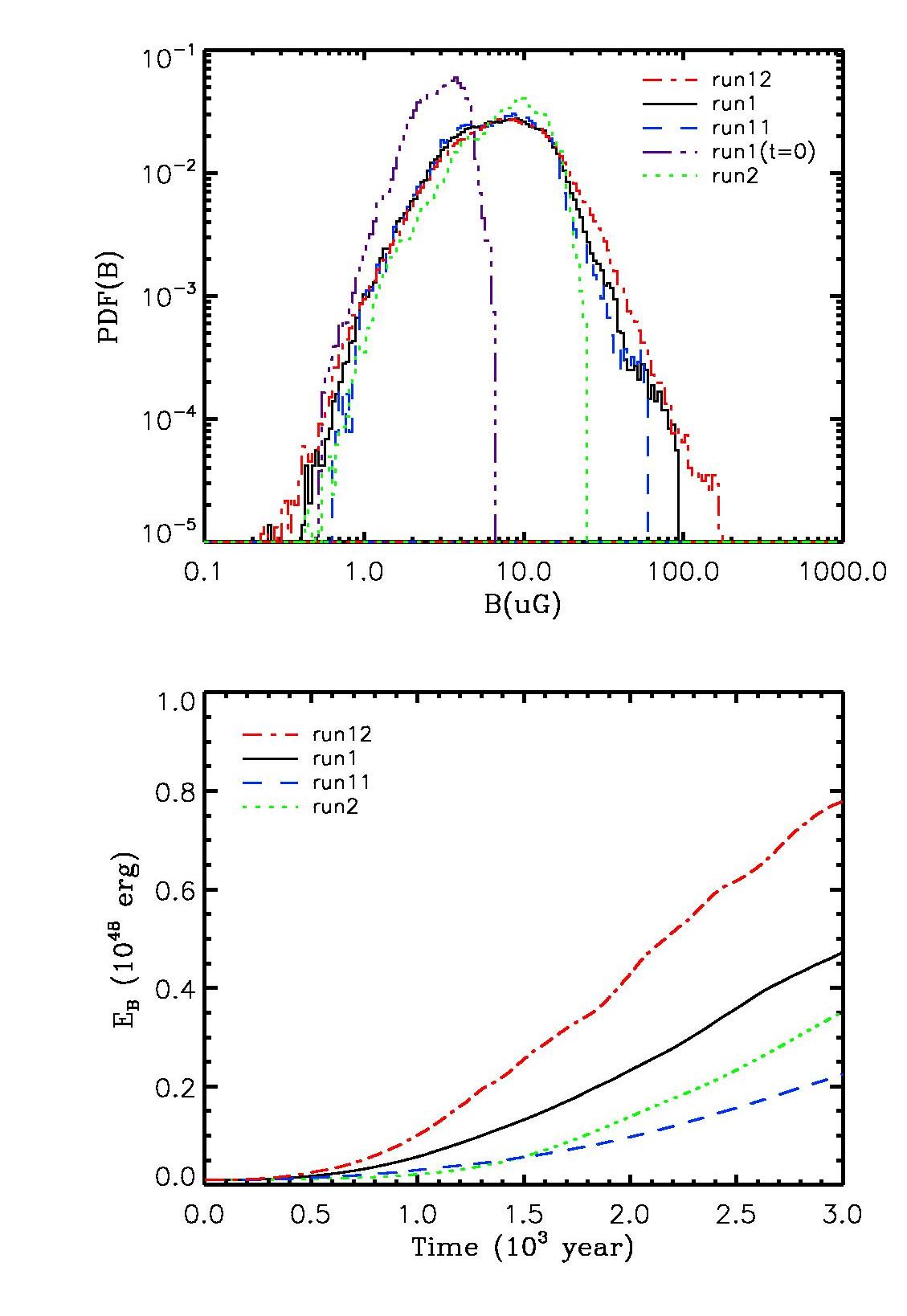}
\caption{(a) The probability distribution function of magnitude
of magnetic field downstream of a distance within $0.3$ pc
to the shock front at $t = 600$ years for runs with different resolution.
The PDF of the background magnetic field in the same region
taken from Run 1 at $t=0$ is also plotted.
(b) The comparison of
total magnetic-field energy evolution for the same runs.
}
\label{fig6}
\end{center}
\end{figure}

Since the field amplification at the shock front is closely related to the vorticity generation,
the numerical resolution is expected to play an important role. Numerical modeling of this process
must resolve the vortical motions.
In Figure \ref{fig6} (a) we plot the normalized probability distribution function (PDF) of
the magnitude of magnetic field downstream within a distance of $0.3$ pc of the shock front
for three different grid resolutions.
Run 1 and 2 have $4000 \times 4000$, Run 11 and 12 have
$2000 \times 2000$, and $8000 \times 8000$, respectively. Figure \ref{fig6} (b) shows the evolution
of the total magnetic-field energy for these cases. It can be seen that higher resolution runs give
higher total magnetic-field energy and, even for the highest resolution of Run 12, the total
amount of magnetic energy in the simulation domain has not converged. More detailed analysis shows that the kinetic energy and magnetic energy have not reached equal partition in small scales. Note that the total
magnetic energy at the end of the simulation ($\sim 3\times 10^3$ yrs) is still a small fraction ($< 10^{-3}$)
of the injected explosion energy or the available kinetic energy in the simulation.

In Figure \ref{fig6} (a), by plotting the PDFs within a small downstream region close to the shock front,
we can get a more clear view of the effects of the numerical resolution in the
shock amplification process near the shock front. The green curve from Run 2 represents
the shock amplification without the background ISM density turbulence, whereas Run 1, 11, and 12 represent
the shock amplification being significantly enhanced when the background ISM turbulence is present.
The purple curve represents the initial, un-shocked background
ISM magnetic-field distribution in the same region as shown in the cases that include ISM density turbulence.
The higher resolution Run 12 obviously produces a greater volume fraction with higher magnetic field and
a larger maximum magnetic field in the downstream region of the shock, although one might
argue that the difference among these runs is not very large, at least for the region within $0.3$ pc of the
shock at this particular time.

In Figure \ref{fig6} (b), however, the difference among different resolutions seems to be much bigger. Such differences
are mostly caused by the region where RTI amplifies the magnetic field (see Fig. \ref{fig3} (b) for
field distribution),  because the total magnetic energy is a summation of both the shock-amplified region and the RTI region.
This is consistent with some previous studies \citep{Jun1995} where numerical resolutions
are shown to be important as well. Such requirements for high resolution might also explain
the previous finding that magnetic-field amplification is not as strong \citep{Balsara2001}  when it is difficult to
employ very high numerical resolution in three-dimensional MHD simulations. We have also checked the long term evolution (up to $3\times10^4$ years) of SNRs similar to \citet{Balsara2001} using two-dimensional high-resolution simulation and we consistently find strong magnetic
field larger than $100 \mu G$ in the downstream region.

\begin{figure}
\centering
\begin{tabular}{cc}
\hfill
\epsfig{file=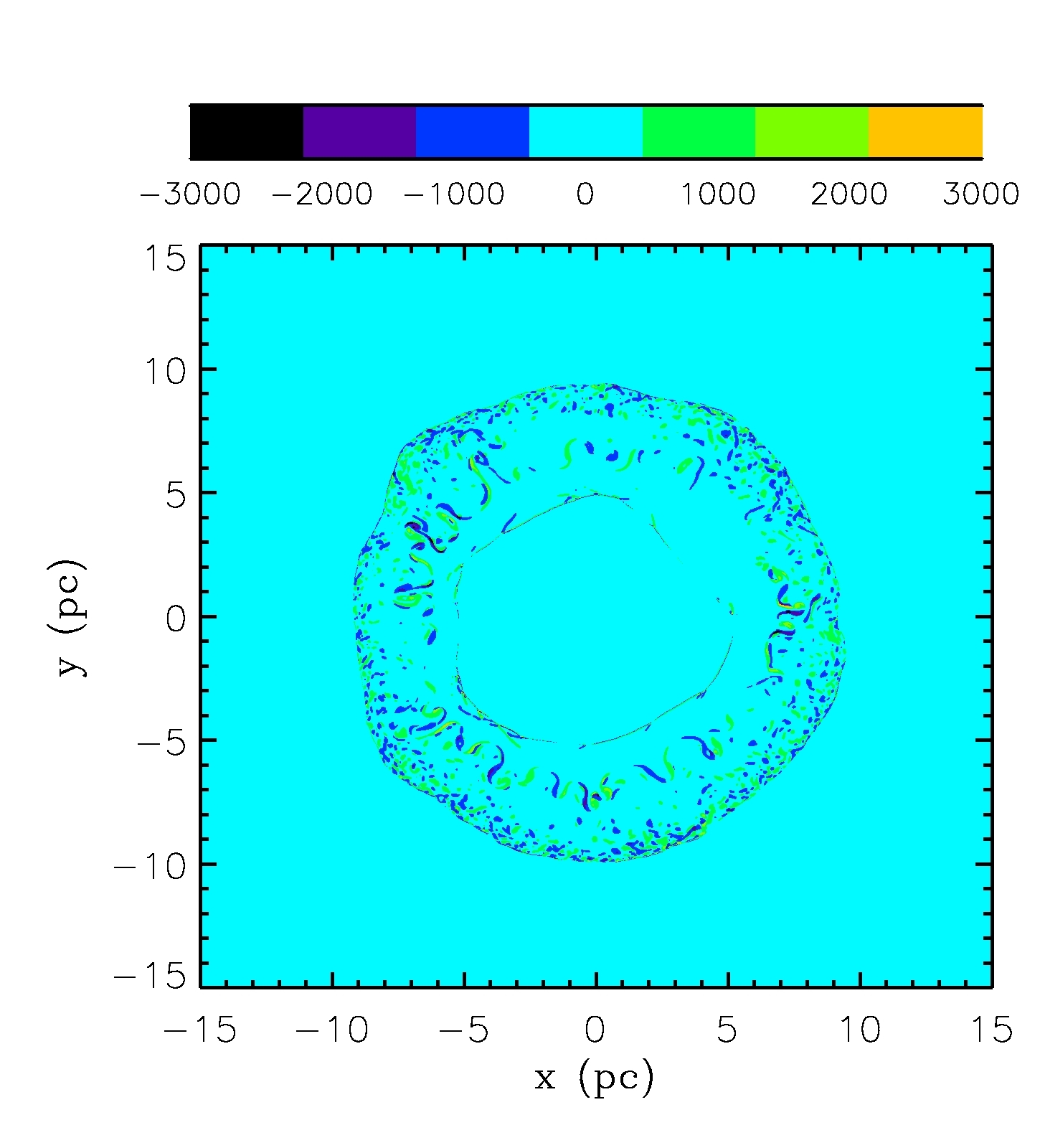,width=0.4\textwidth,clip=} &
\epsfig{file=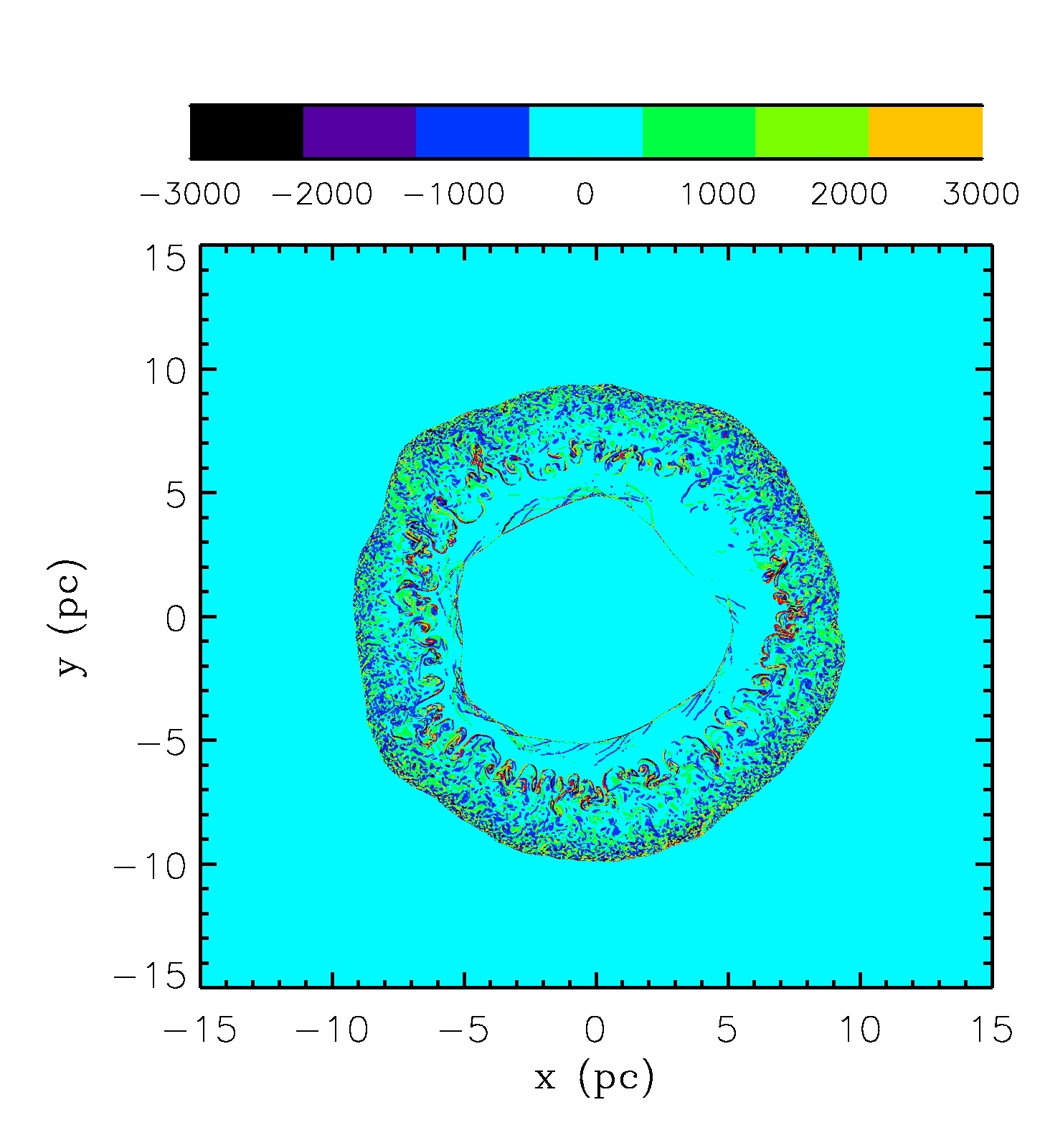,width=0.4\textwidth,clip=} \\
\epsfig{file=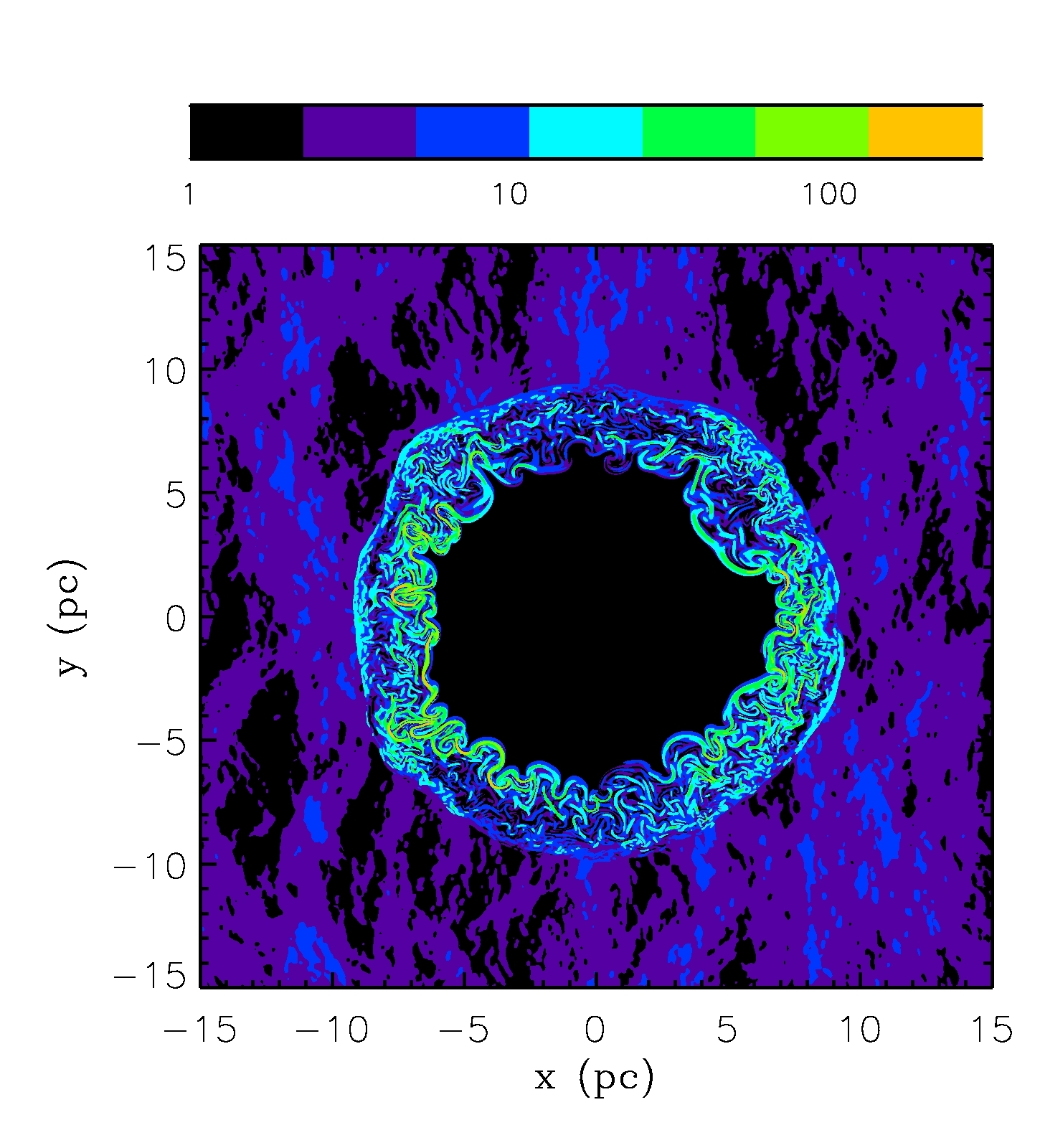,width=0.4\textwidth,clip=} &
\epsfig{file=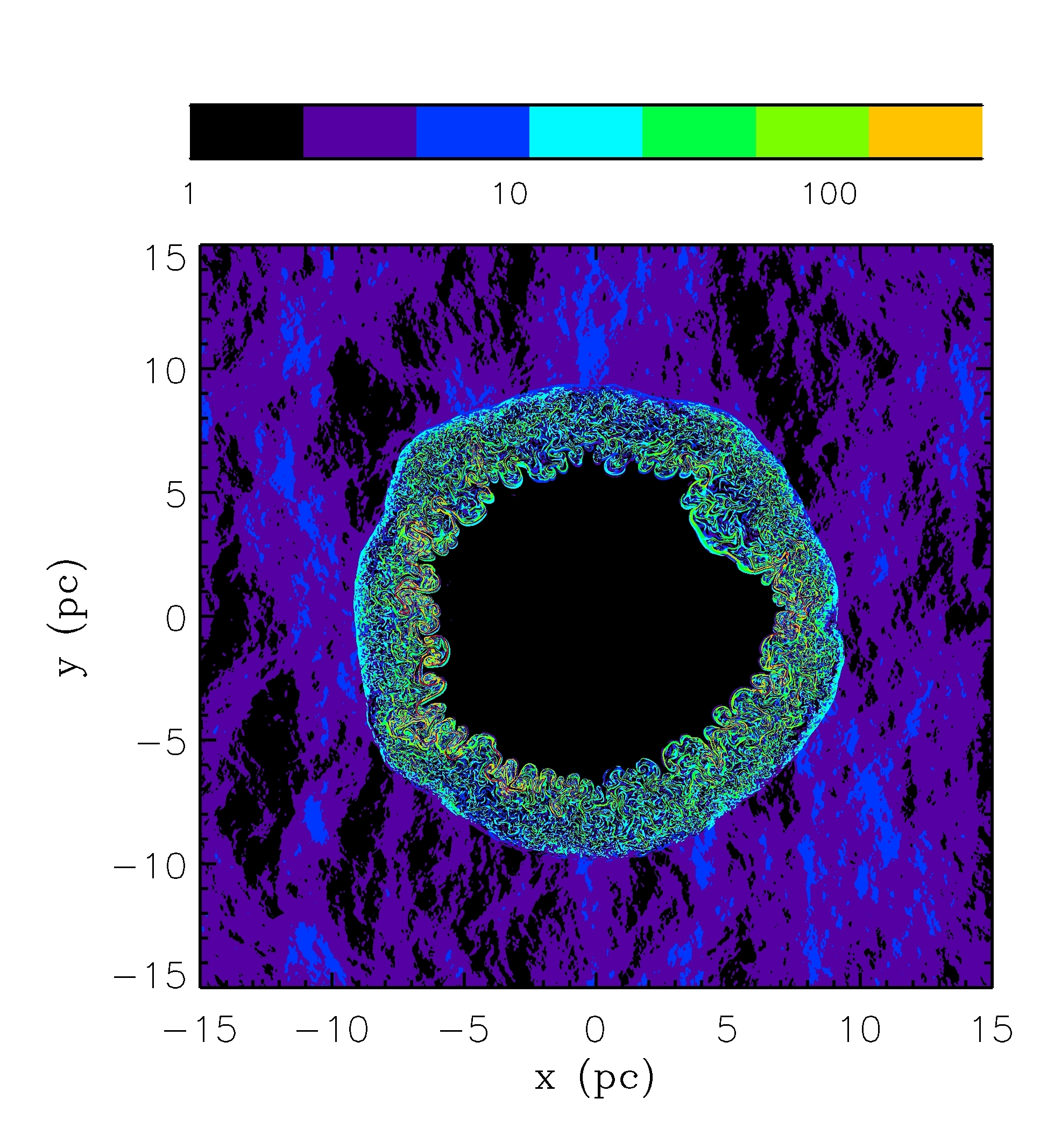,width=0.4\textwidth,clip=}
\end{tabular}
\caption{The color-coded images of vorticity ({\it top}) and the magnetic-field amplitude ({\it bottom})
for Run 11 (low resolution $2000\times 2000$, left panels)
and Run 12 (high resolution $8000\times 8000$, right panels) at $t = 1600$ years.}
\label{fig7}
\end{figure}

Figure \ref{fig6} demonstrates that both the shock and RTI amplification of the magnetic field
depends on the numerical resolution (perhaps more strongly for the RTI region).
Figure \ref{fig7} further supports this conclusion where we plot the vorticity distribution of the shocked
flows for the
low resolution (Run 11) and the high resolution (Run 12), at a time that is the same as in Figure \ref{fig3}.
It is not surprising to see that more small scale structures are developed with much larger vorticity magnitudes
in Run 12. Note that the two magnetic-field amplification regions can be approximately separated spatially (at least at this time).
The small scale structures in vorticity (signifying turbulence) are developed in the immediate downstream of
the forward shock region whereas both relatively large (i.e., the density ``fingers'') and small scale vorticity
features are produced in the RTI region.

\subsection{Other Effects}

\begin{figure}
\begin{center}
\hfill
\includegraphics[width=80mm]{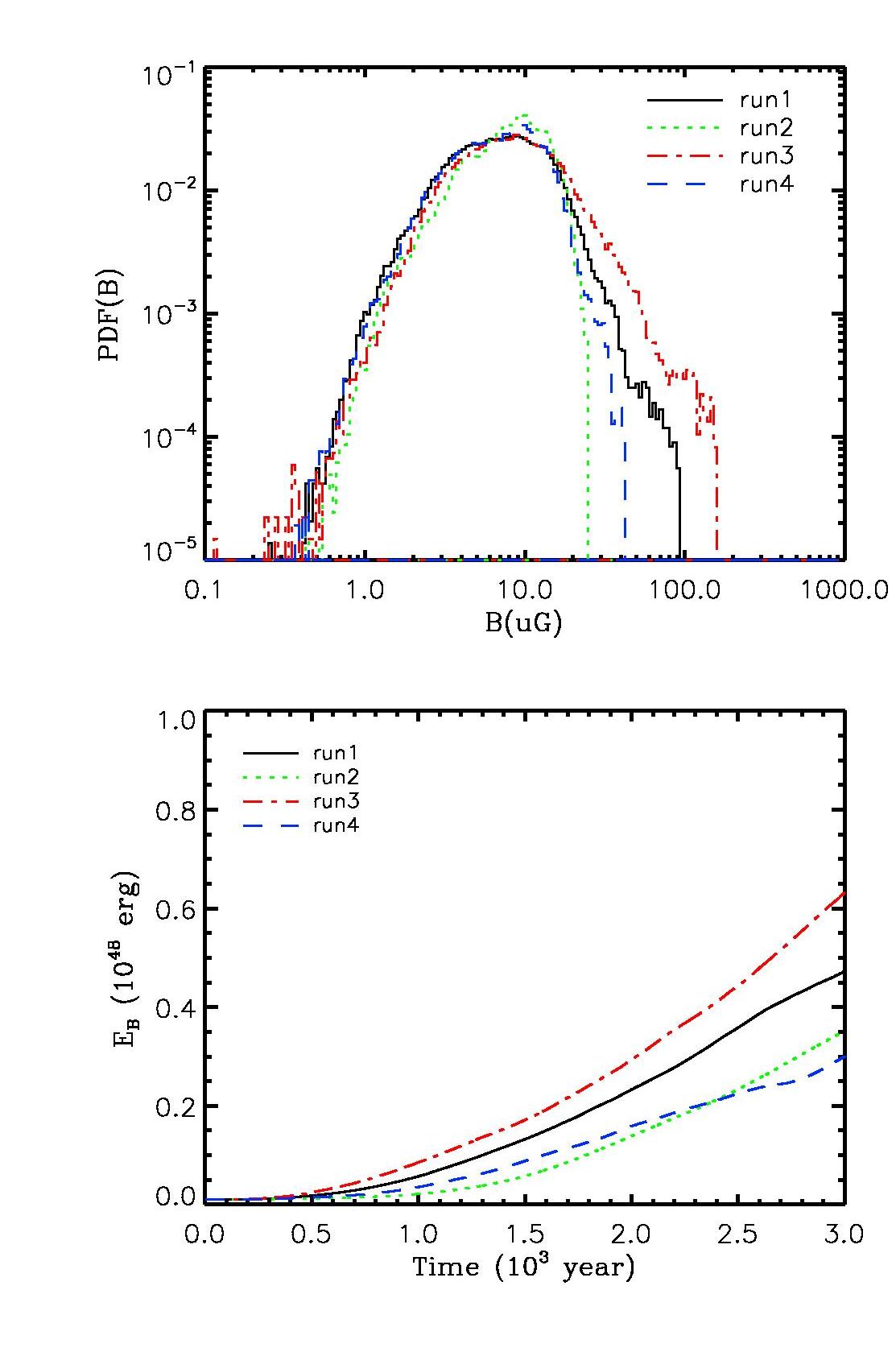}
\caption{Comparison for different background turbulence amplitudes.
(a) The probability distribution function of magnitude
of magnetic field downstream within a distance of
$0.3$ pc behind the shock front at $t = 600$ years. (b) The
comparison of the total magnetic-field energy evolution for these runs.
}
\label{fig8}
\end{center}
\end{figure}

Figure \ref{fig8} shows the effects of different background turbulence amplitudes. In
Figure \ref{fig8} (a), the PDFs of magnetic field within
0.3 pc behind the shock front for Runs 1 - 4  at $t = 600$ years are shown, respectively.
Figure \ref{fig8} (b) represents the evolution of magnetic-field energy for these cases.
It is seen that the larger amplitude density fluctuation tends to lead to
stronger magnetic-field amplification. For Run 2, it can be seen that the effect of
RTI strongly enhance the magnetic field. The total magnetic-field energy can
even exceed the case of Run 4 ($\sqrt{\delta n^2/n_0^2} = 0.3$). Detailed
analysis shows the growth of RTI is somewhat suppressed by turbulence which makes
the magnetic-field energy in Run 4 less than Run 2 at late times.

We also examine the effect of the magnitude of initial interstellar magnetic field
in Run 5 and Run 6. Comparing these two cases with Run 1, we find the shock amplified magnetic field
is nearly proportional to the magnitude of initial magnetic field. This is consistent with the fact that
the amplified fields have not reached saturation.
The effect of the temperature of background medium is also examined in Run 7 and Run 8.
We find that the different temperatures do not yield any strong difference
in the downstream magnetic-field evolution.

\begin{figure}
\begin{center}
\hfill
\includegraphics[width=80mm]{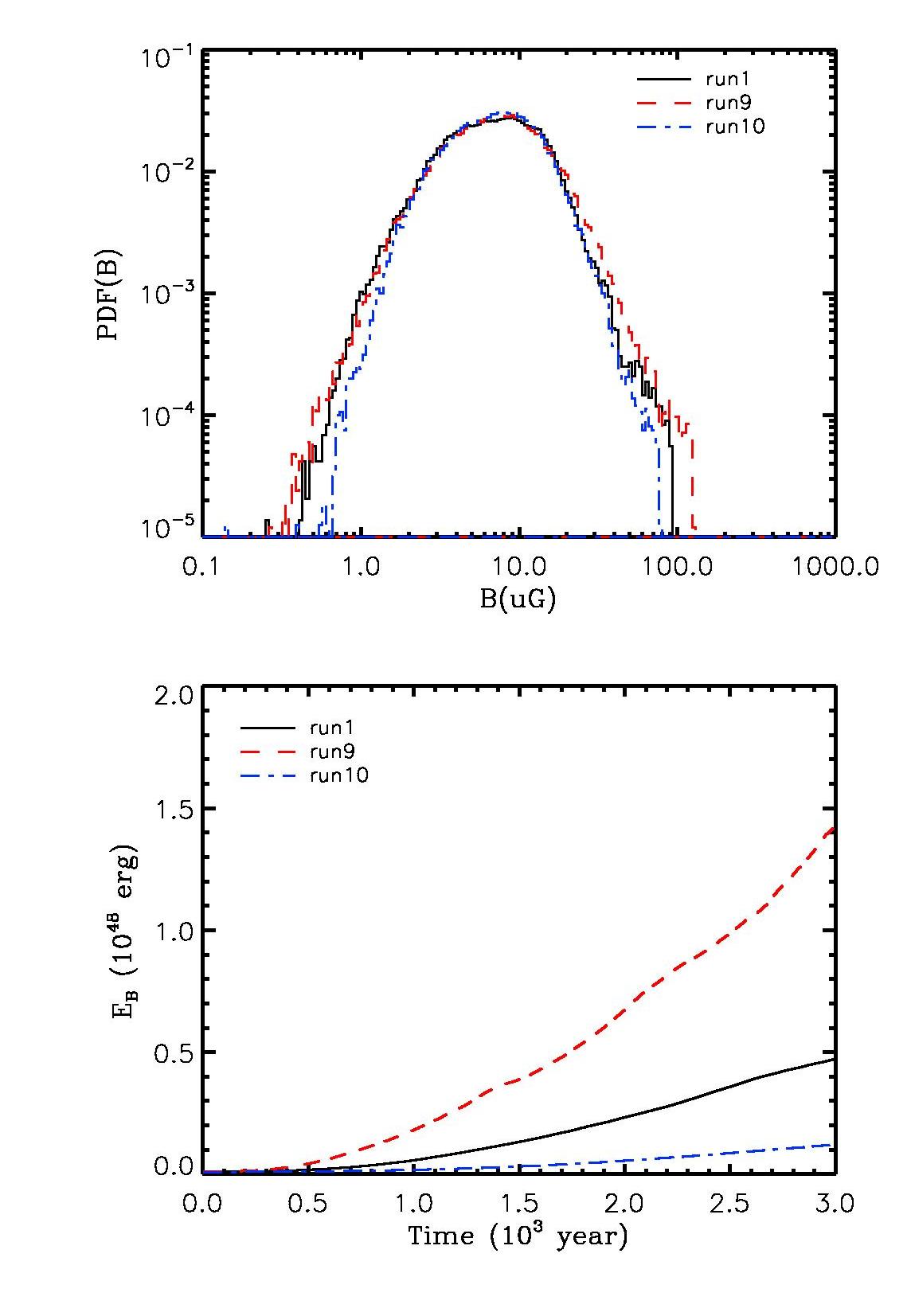}
\caption{Comparison for different initial explosion energies.
(a) The probability distribution function of magnitude
of magnetic field downstream with a distance within $0.3$ pc behind
the shock front when the shock radii are roughly the same for these cases.
(b) Comparison of the total magnetic-field energy for these runs. }
\label{fig9}
\end{center}
\end{figure}

Figure \ref{fig9} is similar to Figure \ref{fig8}, but for different initial
explosion energies, $E = 1.5 \times 10^{51}$ erg, $4.5 \times 10^{51}$ erg,
and $4.5 \times 10^{50}$ erg, respectively. The time
frames are chosen so the shock radii are roughly the same, the
magnetic-field distribution before amplification is therefore
roughly the same. We can see the magnetic-field amplification is
stronger for the case of higher explosion energy. This is because the central
region drives a stronger shock which generates stronger vorticity in the downstream region.
The magnetic energy evolution in these three cases follows the same trend.

\section{Discussion and Conclusion}

The inferred strong magnetic field in young SNRs is a significant result
and can be important in the high-energy processes including particle acceleration
and thermal/nonthermal emissions. The origin of this
process, however, is still under debate. In this work we study the interaction between a
supernova blast wave with a turbulent upstream medium
which contains density and magnetic-field fluctuations. The vorticity
produced at the rippled
shock front can stretch and distort the magnetic field lines, and this leads to
a strong magnetic-field
amplification downstream \citep{Giacalone2007,Inoue2009}. Using two-dimensional
MHD simulations of a blast wave,
we confirm the key features of this process.
Based on our simulations, we conclude that the increase of magnetic field is
dependent on shock speed and background density turbulence amplitudes.
Furthermore, the numerical resolution used in the simulations can play an important role as well.
Previous work \citep{Balsara2001} using three-dimensional MHD simulation with
moderate resolution shows no magnetic-field amplification beyond $50\mu G$.
Here we show the magnetic evolution downstream is sensitive to the resolutions
used in the simulation. For high resolutions, the simulations allow rapid growth
at small scales, this leads to efficient field amplification.

Furthermore, we find that there are two different processes and spatial regions where
magnetic field is amplified. One is associated with the shock amplification immediately downstream
and the other is associated with the RTI at the interface between the ejecta and the shocked medium.

However, in our simulations, we did not observe a systematic strong magnetic field
within a thin region immediate downstream of the supernova shock. For example, using the results
of the highest resolution case, within $0.15$ pc downstream of supernova shock, we
observe only about $0.8\%$ region which has magnetic field larger than $30 \mu G$. This lack
of strong magnetic field can be understood as the downstream dynamo process requires
an efficient stretching to produce strong magnetic field. The time scale for the growth
of magnetic field depends on the eddy turnover time. Only after a certain time can the
field get sufficient amplification. If the thin rims ($0.01 - 0.1$ pc) observed in young SNRs are indeed caused
by the electrons losing energy in strong fields ($\sim$ several hundred $\mu G$),
some other processes such as three-dimensional instabilities, plasma kinetics or the effect of cosmic rays might be
needed to explain the magnetic-field amplification in young SNRs.

We note that in observation there is no one-to-one correspondence between X-ray ``thin rims" and
inferred shock locations. In fact many ``thin rims" and filamentary structures seen in X-ray observation
can hardly be related to shock front due to observation limitation. Further understanding about the relationship
between the small scale X-ray structure and shock locations is needed to further constraint and distinguish
the different mechanisms for amplification of magnetic field.

We also note that the two-dimensional simulation in this study could be significantly different from
three-dimensional simulation.
It is known that in two-dimensional simulation the dynamics of MHD flow is very different from that
for three-dimensional simulation. For example the inverse cascade of enstrophy can causes strong
intermittency due to two-dimensional effect \citep{Biskamp2003}. There are several recent studies
show that magnetic field amplification behind high-mach number shock \citep{Inoue2011a,Inoue2011b} and
in Rayleigh-Taylor region \citep{Stone2007} can still operate in three-dimensions, which confirm
the results found in two-dimensional simulations. Also, in three-dimensional simulation the MHD flow could
develop other types of instabilities with larger growth rates.
Further three-dimensional MHD simulation with high resolution will be
useful in confirming the conclusions of this paper.

\section*{Acknowledgement}
This work was supported by the LDRD and IGPP programs at LANL and by DOE office of science via CMSO.
Computations were performed using the institutional computing resources at LANL. JG, JRJ, and FG also acknowledge partial support from NASA grant NNX10AF24G. JRJ is also supported in part by NASA grant NNX11AB45G.
We acknowledge useful discussions with Dr. Hao Xu and Dr. Federico Fraschetti.

\clearpage
\small

\begin{table}
\begin{tabular*}
{1.\textwidth}{lcccccccc}
\hline
Run&$n$& $\sqrt{\delta n^2/n_0^2}$ & $B_0$ & $T_0$ & $E_{in} (10^{51})$ & Grids & $B_{ave}$(rim) & $B_{max}$(rim)\\
\hline
1  & 1.0 &0.45 & 3& $1.0\times 10^4$ & $1.5 $ & $4000^2$  & 8.7 & 48.0\\
2  & 1.0 &0.0  & 3& $1.0\times 10^4$ & $1.5 $ & $4000^2$  & 8.0 & 23.2\\
3  & 1.0 &0.7  & 3& $1.0\times 10^4$ & $1.5 $ & $4000^2$  & 9.8 & 61.8\\
4  & 1.0 &0.30 & 3& $1.0\times 10^4$ & $1.5 $ & $4000^2$  & 8.3 & 27.5\\
5  & 1.0 &0.45 & 1& $1.0\times 10^4$ & $1.5 $ & $4000^2$  & 2.9 & 16.1\\
6  & 1.0 &0.45 & 9& $1.0\times 10^4$ & $1.5 $ & $4000^2$  &25.9 & 143.4\\
7  & 1.0 &0.45 & 3& $1.0\times 10^5$ & $1.5 $ & $4000^2$  & 8.7 & 48.1\\
8  & 1.0 &0.45 & 3& $1.0\times 10^6$ & $1.5 $ & $4000^2$  & 8.7 & 48.2\\
9  & 1.0 &0.45 & 3& $1.0\times 10^4$ & $4.5 $ & $4000^2$  & 9.0 & 63.2\\
10 & 1.0 &0.45 & 3& $1.0\times 10^4$ & $0.45 $ & $4000^2$ & 8.3 & 37.7\\
11 & 1.0 &0.45 & 3& $1.0\times 10^4$ & $1.5 $ & $2000^2$  & 8.3 & 35.6\\
12 & 1.0 &0.45 & 3& $1.0\times 10^4$ & $1.5 $ & $8000^2$  & 9.1 & 72.8\\
 \hline
\end{tabular*}
 \caption{The parameters used in the simulations. Number density $n$ is in cm$^{-3}$, magnetic field $B$ is in $\mu G$, temperature $T_0$ is in $K$, and explosion energy $E$ is in ergs. The last two columns list the average and maximum magnetic-field strength in unites of $\mu G$ within a thin region (0.15 pc) downstream of supernova shock at 600 years. In all the cases, the fluctuating component of the magnetic field is taken to be $\delta B^2 = B_0^2$}
\end{table}

\clearpage

\end{document}